\documentclass[sigconf]{acmart}
\AtBeginDocument{%
  }
\usepackage{multirow} 
\usepackage{pifont}
\usepackage{algpseudocode}
\usepackage{amsmath}
\usepackage{xcolor}

\usepackage{graphicx}  
\usepackage{adjustbox} 
\usepackage{xcolor} 
\usepackage{url}    

\usepackage{multirow}
\usepackage{booktabs}
\usepackage{amsmath}

\usepackage{soul}
\usepackage{url}
\usepackage{hyperref}

\usepackage{graphicx}
\usepackage{subcaption}
\usepackage{caption}
\usepackage{adjustbox}

\usepackage{xcolor}
\usepackage[table]{xcolor}
\usepackage{color}
\usepackage{colortbl}

\usepackage{booktabs}
\usepackage{multirow}
\usepackage{amsmath}

\usepackage{amsthm}

\usepackage{algorithm}

\usepackage[most]{tcolorbox}
\usepackage[switch]{lineno}

\renewcommand\footnotetextcopyrightpermission[1]{}
\definecolor{softgreen}{RGB}{0,154,85} %
\definecolor{softred}{RGB}{250,120,130}

\newcommand{\eg}{\textit{e.g.}}
\newcommand{\ie}{\textit{i.e.}}
\newcommand{\vs}{\textit{vs.}~}
\newcommand{\etal}{\textit{et al.}}

\setcopyright{acmlicensed}
\copyrightyear{2026}
\acmYear{2026}
\acmDOI{XXXXXXX.XXXXXXX}
\acmConference[arXiv Preprint]{arXiv Preprint}{March, 2026}{arXiv.org}

\begin{document}

\title{MAR3: Multi-Agent Recognition, Reasoning, and Reflection for Reference Audio-Visual Segmentation}




\author{Yuan Zhao}
\affiliation{%
  \department{College of Computer Science}
  \institution{Inner Mongolia University}
  \city{Hohhot}
  \country{China}}
\email{zy404nf@163.com}

\author{Zhenqi Jia}
\affiliation{%
  \department{College of Computer Science}
  \institution{Inner Mongolia University}
  \city{Hohhot}
  \country{China}}
\email{jiazhenqi7@163.com}

\author{Yongqiang Zhang*}
\affiliation{%
  \department{College of Computer Science}
  \institution{Inner Mongolia University}
  \city{Hohhot}
  \country{China}}
\email{zhangyongqiang@imu.edu.cn}

\renewcommand{\shortauthors}{Zhao \etal}

\begin{abstract}
Reference Audio-Visual Segmentation (Ref-AVS) aims to segment objects in audible videos based on multimodal cues in reference expressions. Previous methods overlook the explicit recognition of expression difficulty and dominant modality in multimodal cues, over-rely on the quality of the instruction-tuning dataset for object reasoning, and lack reflective validation of segmentation results, leading to erroneous mask predictions. To address these issues, in this paper, we propose a novel training-free \textbf{M}ulti-\textbf{A}gent \textbf{R}ecognition, \textbf{R}easoning, and \textbf{R}eflection framework to achieve high-quality Reference Audio-Visual Segmentation, termed \textbf{MAR3}. Incorporating the sociological Delphi theory to achieve robust analysis, a Consensus Multimodal Recognition mechanism is proposed that enables LLM agents to explicitly recognize the difficulty of reference expressions and the dominant modality of multimodal cues. Based on our modality-dominant difficulty rule, we propose an adaptive Collaborative Object Reasoning strategy to reliably reason about the referred object. To further ensure precise mask prediction, we develop a Reflective Learning Segmentation mechanism, in which a check agent examines intermediate segmentation results and iteratively corrects the object text prompt of the segment agent. Experiments demonstrate that MAR3 achieves superior performance (69.2\% in $\mathcal{J}\&\mathcal{F}$) on the Ref-AVSBench dataset, outperforming SOTA by 3.4\% absolutely. 

\end{abstract}

\keywords{Reference Audio-Visual Segmentation, Multimodal Reasoning and Understanding, Multi-Agent Systems}





\maketitle

\begin{figure}[ht]
    \centering
    \includegraphics[width=1.0\linewidth]{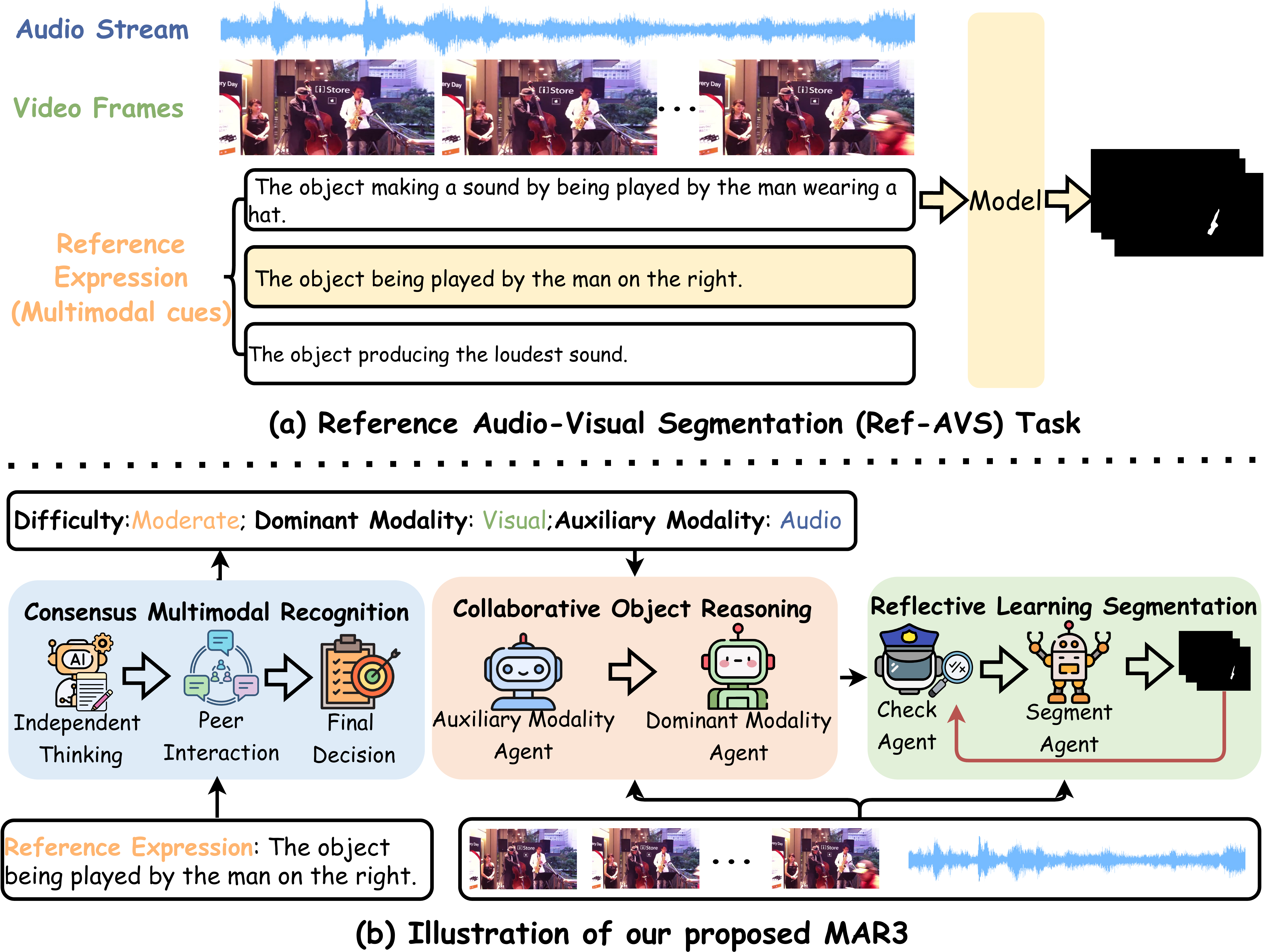}

    \caption{(a) Illustration of the Reference Audio-Visual Segmentation (Ref-AVS) task. (b) Our proposed MAR3 first explicitly recognizes the difficulty of reference expressions and the dominant modality of multimodal cues, then performs referred object reasoning based on the modality-dominant difficulty rule, and finally generates the precise prediction mask through iterative optimization of the referred object prompt based on intermediate segmentation results. }
    \label{fig_1}
\end{figure}

\section{Introduction}
Inspired by the fact that humans perceive the world through multiple sensory organs, multimodal audio-visual reasoning and understanding (\eg distinguishing objects of interest) have attracted growing attention in the AI community. 
As a novel task,
Reference Audio-Visual Segmentation (Ref-AVS) \cite{EMCC} aims to segment objects in audible videos based on multimodal cues (\ie audio and visual) in the reference expression, as shown in Figure \ref{fig_1}(a). It has substantial potential in real-world applications such as film production \cite{EMCC}. In complex dynamic audiovisual scenes, the need for effective multimodal information integration, deep understanding of natural language references, and precise pixel-level segmentation makes the Ref-AVS task particularly challenging.

Existing Ref-AVS research can be broadly categorized into two groups. The first group of methods \cite{EMCC,TSAM-CVPR,SAM2LOVE} focuses on integrating audio, video and text into unified multimodal features, which are then used as prompts to guide segmentation decoders \cite{SAM,SAM2} in generating binary masks of the referred object. However, these methods typically perform only implicit processing of multimodal input, overlooking the cues contained in the reference expression that relate to the contributions of the audio and video modalities. For example, as shown in Figure \ref{fig_1}(a), for the reference expression of \textquotedblleft The object being played by the man on the right.\textquotedblright, we can analyze that the primary focus should be on the visual modality, while the audio modality can serve as an auxiliary, and the difficulty is relatively moderate. However, this aspect is often overlooked in previous research. 

The other family of methods \cite{Crab,TGS} first tunes a multimodal large language model (MLLM) for referred object reasoning, and then proceeds to segmentation. These methods typically rely on multimodal large language models (\eg Gemini-1.5-Pro \cite{gemini_pro}) that have more parameters than the model itself and construct high-quality instruction-tuning datasets through a carefully designed chain of thought. However, not only does this require significant computational resources, but, more importantly, these methods are heavily dependent on the quality of the instruction-tuning dataset, which can result in suboptimal reasoning performance for the referred objects. Furthermore, existing methods generally use a single-stage segmentation process, without further verification to ensure that the segmentation results match the reference expression, leading to inaccurate segmentation mask predictions.

To address the above issues, we propose a novel training-free \textbf{M}ulti-\textbf{A}gent \textbf{R}ecognition, \textbf{R}easoning, and \textbf{R}eflection framework to achieve high-quality Reference Audio-Visual Segmentation, termed \textbf{MAR3}  (as shown in Figure \ref{fig_1}(b)). It contains three mechanisms: (1) Drawing inspiration from human cognitive processes, where task instruction analysis precedes execution, we design a Consensus Multimodal Recognition (CMR) mechanism that enables different LLM agents to explicitly recognize the difficulty of reference expressions and the dominant modality of the multimodal cues. In particular, we incorporate the sociological Delphi theory \cite{linstone1975delphi,Delphi_2,Delphi_3,Delphi_4,Delphi_5,Delphi_6}---a structured decision-making framework originally developed for expert consensus forecasting---to achieve robust analysis results. Different Multimodal Cues Analysis (MCA) agents first independently think about the reference expression, then share and cross-reference their results and reason during peer interaction to reach a more unbiased understanding, and finally one final MCA agent consolidates the discussion results and makes the final decision. (2) To reliably reason about the referred object, we propose an adaptive Collaborative Object Reasoning strategy. According to our defined modality-dominant difficulty rule, for low-difficulty expressions, the dominant-modality agent performs reasoning independently; for moderate- or high-difficulty cases, we adopt a multi-agent collaboration approach, where the dominant-modality agent integrates relevant information from auxiliary-modality agents (\eg the candidate object list and reasoning rationale) to derive a more precise result.
(3) We propose a Reflective Learning Segmentation mechanism, which incorporates a check agent that examines intermediate segmentation results to verify their consistency with the reference expression, and then iteratively refines the object text prompts of the segment agent based on this feedback. Leveraging such reflective learning, our method effectively improves the accuracy of mask prediction.

In summary, the main contributions of this paper can be summarized as follows:

\begin{itemize}
\item We propose \textbf{MAR3}, a training-free multi-agent framework that decomposes the Ref-AVS task into phases of Recognition, Reasoning, and Reflection, enabling a deep understanding of multimodal cues, reliable reasoning about the referred objects, and precise mask prediction.
\item We design a \textit{Consensus Multimodal Recognition} mechanism to explicitly recognize the difficulty and dominant modality in the reference expression and multimodal cues using the sociological Delphi theory. An adaptive \textit{Collaborative Object Reasoning} strategy is proposed to reliably reason about the referred object by establishing collaboration between dominant- and auxiliary-modality agents. Additionally, we develop a \textit{Reflective Learning Segmentation} mechanism to ensure precise mask prediction where a check agent iteratively corrects the object text prompts for the segment agent based on intermediate segmentation results.
\item Extensive experiments demonstrate that our proposed MAR3 achieves superior performance (69.2\% in $\mathcal{J}\&\mathcal{F}$) on the Ref-AVS benchmark, outperforming SOTA by 3.4\%  absolutely, while various ablation studies validate the effectiveness of each proposed module.
\end{itemize}

\begin{figure*}[!t]
    \centering
    \includegraphics[width=0.956\linewidth]{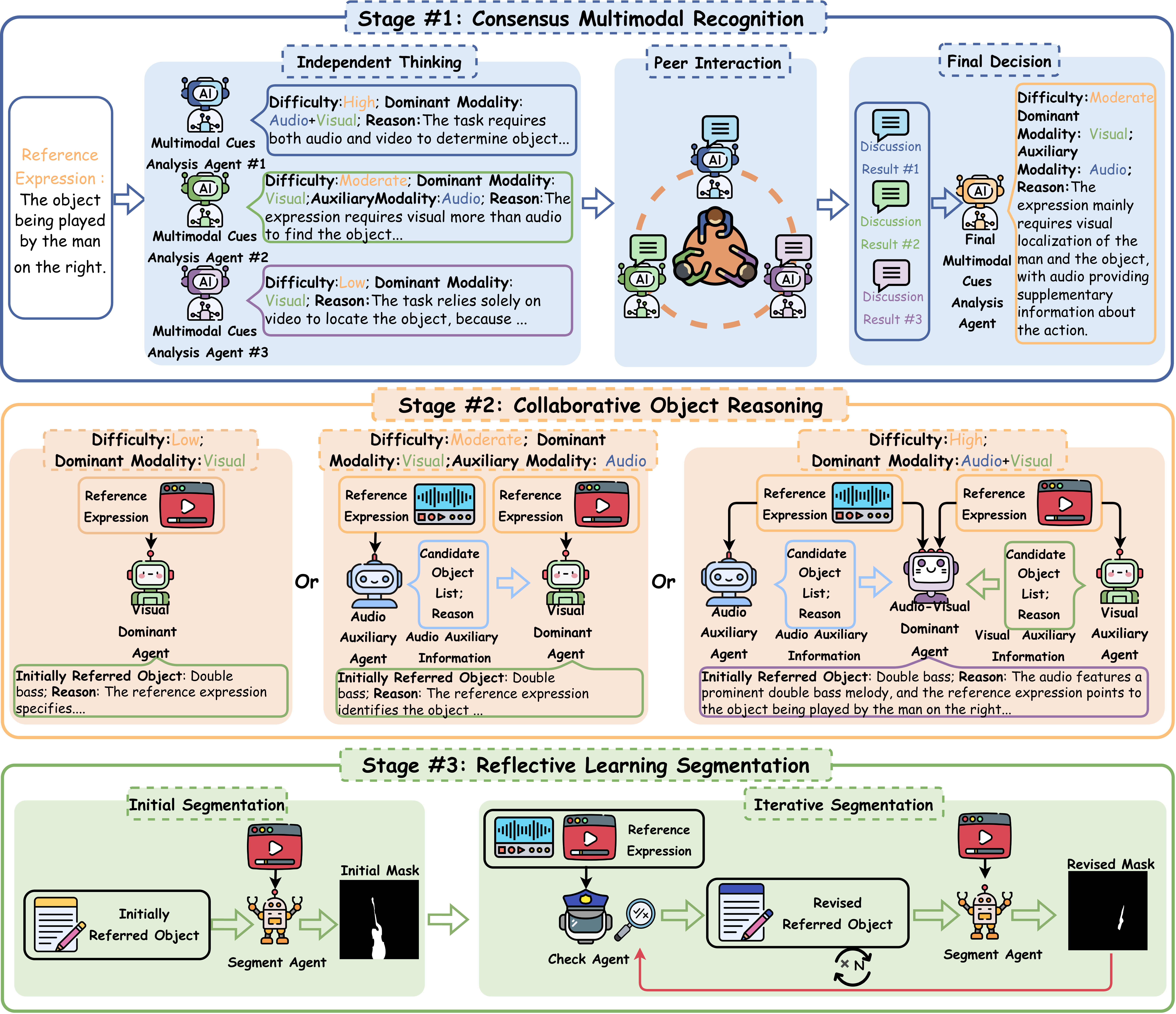}
    \caption{The architecture of our MAR3, which contains three mechanisms: Consensus Multimodal Recognition (CMR), Collaborative Object Reasoning (COR), and Reflective Learning Segmentation (RLS). CMR explicitly recognizes the difficulty and dominant modality in the referring expression and multimodal cues using the sociological Delphi theory. COR is designed to reliably reason about the referred object through collaboration between dominant- and auxiliary-modality agents. RLS ensures precise mask prediction where a check agent iteratively corrects the object text prompts for the segment agent based on intermediate segmentation results.
    }
    \label{fig_2}
\end{figure*}

\section{Related Works}

\subsection{Reference Audio-Visual Segmentation}
Reference Audio-Visual Segmentation (Ref-AVS) \cite{EMCC} aims to generate binary masks of objects in audible videos based on multimodal cues (\ie audio and visual) in the reference expression. Some works \cite{EMCC,TSAM-CVPR,SAM2LOVE} focus on integrating audio, video, and text into unified multimodal features and then using them as prompts to guide the segmentation decoder \cite{SAM,SAM2} to generate precise masks across video frames. For example, SAM2-LOVE \cite{SAM2LOVE} designs a multimodal fusion Transformer to integrate textual, audio, and visual information into a learnable \texttt{[seg]} token, which serves as a prompt for the fine-tuned SAM2 \cite{SAM2} decoder to perform final video segmentation. In addition, some works \cite{Crab,TGS} first instruction-tune a multimodal large language model (MLLM) to reason about the target object referred by the expression, and then perform the subsequent segmentation stage. For instance, TGS-Agent \cite{TGS} trains a reasoning-enhanced MLLM using an instruction-tuning dataset, which converts multimodal inputs into descriptions of the referred object, then uses it to guide Grounding-DINO \cite{groundingdino} and SAM2 \cite{SAM2} for localization and segmentation. However, existing methods neglect explicit recognition of expression difficulty and dominant modality, both of which are crucial for Ref-AVS. They also over-rely on the quality of the instruction-tuning dataset, resulting in suboptimal object reasoning, and lack reflective checks, causing erroneous mask predictions.

To address the above-mentioned issues, a Consensus Multimodal Recognition mechanism is proposed to recognize the difficulty and dominant modality in the reference expression and multimodal cues. We propose a Collaborative Object Reasoning strategy to reliably reason about the referred object, and a Reflective Learning Segmentation mechanism to ensure precise mask prediction.

\subsection{Multi-Agent System}

Recent advances in multimodal large language models (MLLMs) have boosted the development of artificial intelligence multi-agent systems (MAS), which are composed of multiple autonomous or semi-autonomous agents that work together through collaboration, communication, and coordination to complete complex tasks \cite{multi-agent1,multi-agent2, multi_agent3, Hm-rag, Genmac,wang2024macrec}. MAS has been widely applied to a variety of tasks, such as text-to-image generation \cite{chen2025t2i,text-image-2,MCCD-text-to-image}, medical diagnosis \cite{Tree-agent,kim2024mdagents,zhou2025mam}, and audio generation \cite{audiogen1,audiogen2,li2025dialogueagents}. For example, MCCD \cite{MCCD-text-to-image} designs a multi-agent scene parsing module that uses MLLMs to effectively extract detailed scene elements from text, generating realistic and aesthetically pleasing complex scene images. Tree-of-Reasoning \cite{Tree-agent} proposes a multi-agent framework that enhances reasoning ability and interpretability in complex medical diagnosis tasks by introducing a reasoning tree structure based on clinical evidence and a multi-agent cross-validation mechanism. 

In dynamic complex audio-visual scenes, the Ref-AVS task requires effective integration of multimodal information, deep understanding of natural language references, and precise pixel-level segmentation, making it a highly coupled task involving multiple complex objectives. The MAS system provides an ideal solution by decomposing complex problems into coordinated sub-tasks and assigning specialized roles to each task. Therefore, we apply the MAS system to the Ref-AVS task for the first time. Specifically, we decompose the Ref-AVS task into three sub-tasks (\ie Recognition, Reasoning, and Reflection) to achieve high-quality reference audio-visual segmentation. Moreover, our proposed training-free MAR3 framework reduces the need for computationally expensive joint optimization, effectively addresses the core challenges of the Ref-AVS task, and significantly enhances the system’s flexibility and adaptability.

\section{MAR3: Methodology}

\subsection{Overview}
 
In this paper, we propose MAR3, a novel training-free Multi-Agent Recognition, Reasoning, and Reflection framework, tailored for high-quality Reference Audio-Visual Segmentation (Ref-AVS). MAR3 is designed to enable deep multimodal cue comprehension, reliable referred object reasoning, and accurate segmentation mask prediction.
The main architecture of our proposed method is shown in Figure \ref{fig_2}; it contains three mechanisms: (1) In the Consensus Multimodal Recognition (CMR) mechanism, we construct different multimodal cue analysis agents to explicitly recognize the difficulty of reference expressions and the dominant modality of the multimodal cues through independent thinking, peer interaction, and final decision processes. (2) In Collaborative Object Reasoning (COR) mechanism, for low-difficulty expressions, the dominant modality agent performs reasoning independently, while for moderate- and high-difficulty scenarios, a collaborative paradigm is adopted where the dominant modality agent integrates essential information provided by auxiliary modality agents to perform referred object reasoning. (3) In Reflective Learning Segmentation (RLS) mechanism, we devise a check agent that examines intermediate segmentation results and iteratively corrects the object text prompt of the segment agent. 

\subsection{Consensus Multimodal Recognition}

Drawing inspiration from the human cognitive process, where the analysis instruction occurs before executing a task, we design the Consensus Multimodal Recognition (CMR) mechanism to explicitly recognize the difficulty of reference expressions and the dominant modality of multimodal cues. Specifically, 
we first introduce a modality-dominant difficulty rule to correlate the difficulty level of reference expressions with multimodal cues, which is categorized as follows:
(1) Low Difficulty: When the dominant modality is either audio or visual. (2) Moderate Difficulty: When the dominant modality is audio and the auxiliary modality is visual, or vice versa. (3) High Difficulty: When both audio and visual are dominant modalities. 
Then, to achieve a deep understanding of multimodal cues, we introduce the sociological Delphi Method \cite{linstone1975delphi,Delphi_2,Delphi_3,Delphi_4,Delphi_5,Delphi_6} into our recognition process. The Delphi method relies on a panel of experts who participate in independent evaluation, anonymous feedback, and final decision-making to reach a high-quality consensus. By mapping this sociological framework to LLM agents, we formulate a systematic recognition process, which includes independent thinking, peer interaction, and a final decision-making phase to yield robust analysis results.
For the output of each phase, we define three key elements: difficulty, dominant and auxiliary modality, and reason.

\noindent
\textbf{Independent Thinking.} To mitigate potential biases inherent in a single-agent architecture and enhance the robustness and accuracy of the analytical outcomes, we employ three heterogeneous LLM agents as Multimodal Cues Analysis (MCA) agents. Each MCA agent independently evaluates the following dimensions of a given reference expression: difficulty, dominant modality, and reason. We apply the predefined modality-dominant difficulty rule to the prompt, where each agent receives the same input and derives its own understanding and conclusions through their independent thinking.

\noindent
\textbf{Peer Interaction.} According to Delphi theory \cite{linstone1975delphi,Delphi_2,Delphi_3,Delphi_4,Delphi_5,Delphi_6}, multiple rounds of expert feedback and interaction can effectively reduce biases. Inspired by this, we design a peer interaction mechanism. In this process, each MCA agent refers to the independent thinking results of the other two MCA agents (including difficulty, dominant modality, and reason), then reanalyzes the reference expression based on these results and generates its own discussion result. The output format is consistent with the initial analysis. Through peer interaction, the agents can validate and correct the analysis results of each other, thus enhancing the overall accuracy and consistency of the analysis.

\noindent
\textbf{Final Decision.} After peer interaction, the final MCA agent makes a decision based on the discussion results of each agent, forming a unified conclusion. By integrating the perspectives of different agents, the final agent gains a more comprehensive understanding of the multimodal cues and the difficulty of the reference expression. This final unified conclusion will then serve as the basis for subsequent reasoning and segmentation processes.

Through Delphi-based consensus mechanism, we can accurately recognize the difficulty of the reference expression and the dominant modality of the multimodal cues, providing more robust support for subsequent reliable reasoning about the referred objects and precise segmentation mask prediction.

\subsection{Collaborative Object Reasoning}

In multi-agent systems, collaboration has become a vital paradigm for solving complex tasks \cite{collaborative_1,collaborative_2,Lvagent,ding2025muses,yang2025nader,ouyang2025nvagent,hu2024storyagent}. Inspired by human team collaboration and collective reasoning processes, we design a Collaborative Object Reasoning (COR) mechanism.  This mechanism enables reliable reasoning about the referred object from dynamic audio-visual scenes by establishing collaboration between dominant modality agents and auxiliary modality agents. 
Specifically, we adjust the reasoning path of agents based on our modality-dominant difficulty rule: for low-difficulty expressions, the dominant modality agent is directly employed, while for moderate- or high-difficulty cases, the dominant modality agent integrates vital information provided by auxiliary-modality agents to perform reasoning.

\noindent
\textbf{Difficulty: Low.} For low-difficulty cases, if the dominant modality is visual, the visual agent receives video frames and the reference expression; if the dominant modality is audio, the audio agent receives audio streams and the reference expression. In both cases, the agent independently reasons about the initially referred object and its reason.

\noindent
\textbf{Difficulty: Moderate.} For moderate-difficulty cases, when the dominant modality is visual, and the auxiliary modality is audio, the audio auxiliary agent receives the audio stream and reference expression and generates the candidate object list and reason, which are passed to the visual dominant agent. The visual dominant agent, based on audio auxiliary information, reference expression, and video frames, independently reasons the initially referred object and its reason. If the dominant modality is audio and the auxiliary modality is visual, a similar reasoning process is executed.

\noindent
\textbf{Difficulty: High.} In high-difficulty cases, where both audio and visual modalities are dominant, the audio auxiliary agent and the visual auxiliary agent independently analyze their respective modalities and reference expressions, and generate a candidate object list and reason. Based on the audio stream, video frames, and reference expression, the audio-visual dominant agent receives and integrates this auxiliary information, jointly reasoning about the initially referred object. This collaborative object reasoning process ensures that both modalities contribute to the final reasoning, enabling reliable reasoning in complex audio-visual scenes.

Through this adaptive collaboration strategy, our MAR3 effectively integrates the strengths of both dominant and auxiliary modality agents, ensuring a reliable foundation for subsequent precise segmentation mask prediction.

\begin{table*}[!ht]
\centering
\small
\setlength{\tabcolsep}{12pt}
\renewcommand{\arraystretch}{1.05}
\caption{Comparison with other methods on Ref-AVSBench dataset. $\mathcal{J}\&\mathcal{F} = (\mathcal{J} + \mathcal{F})/2$ in each subset. ``$\uparrow$'' indicates higher is better. 
(Though the asterisked (*) results do not achieve optimal performance, the performance gap remains minimal.)
}
\label{tab:comparison}
\begin{tabular}{l|c|ccc|ccc|ccc}
\toprule
\multirow{2}{*}{\textbf{Method}} & \multirow{2}{*}{\textbf{Venue}} & \multicolumn{3}{c|}{\textbf{Mix (S+U)} $\uparrow$} & \multicolumn{3}{c|}{\textbf{Seen} $\uparrow$} & \multicolumn{3}{c}{\textbf{Unseen} $\uparrow$} \\
 & & $\mathcal{J}$ & $\mathcal{F}$ & $\mathcal{J}\&\mathcal{F}$ & $\mathcal{J}$ & $\mathcal{F}$ & $\mathcal{J}\&\mathcal{F}$ & $\mathcal{J}$ & $\mathcal{F}$ & $\mathcal{J}\&\mathcal{F}$ \\
\midrule
EEMC \cite{EMCC} & ECCV 2024 & 41.9 & 58.1 & 50.0 & 34.2 & 51.3 & 42.8 & 49.5 & 64.8 & 57.2 \\
TSAM \cite{TSAM-CVPR} & CVPR 2025 & 49.0 & 61.6 & 55.3 & 43.4 & 56.8 & 50.1 & 54.6 & 66.4 & 60.5 \\
SAM2-LOVE \cite{SAM2LOVE} & CVPR 2025 & 55.0 & 62.1 & 58.6 & 43.5 & 51.9 & 47.7 & 66.5 & 72.3 & 69.4 \\
Crab \cite{Crab} & CVPR 2025 & 43.1 &  &  & 40.5 &  &  & 45.6 &  &  \\
OISA \cite{OISA}  & ICCV 2025 & 54.5 & 61.4 & 58.0 & 51.7 & 58.7 & 55.2 & 58.3 & 65.1 & 61.7 \\
TGS-Agent \cite{TGS} & AAAI 2026 & 61.2 & 70.3 & 65.8 & 49.4 & 60.1 & 54.8 & \textbf{72.9} & \textbf{80.4} & \textbf{76.7} \\
\midrule
\textbf{MAR3 (Ours)} & - & \textbf{64.1} & \textbf{74.2} & \textbf{69.2} & \textbf{56.1} & \textbf{68.2} & \textbf{62.2} & 72.1* & 80.2* & 76.2* \\
\bottomrule
\end{tabular}

\end{table*}

\subsection{Reflective Learning Segmentation}

In human cognitive processes, reflective learning is crucial to continuously improving the quality of decision-making. Based on this theory, Reflective Learning Segmentation (RLS) is proposed to achieve accurate segmentation mask prediction through iterative correction of the referred object prompt based on intermediate segmentation results. Specifically, our RLS module is divided into two stages: Initial Segmentation and Iterative Segmentation, details of which are provided in the following sections.

\noindent
\textbf{Initial Segmentation.} In the Initial Segmentation stage, the segment agent receives the initially referred object output from the collaborative object reasoning process as a text prompt and performs segmentation on the video frames to generate an initial prediction mask. 

\noindent
\textbf{Iterative Segmentation.} In the Iterative Segmentation stage, the check agent inspects the audible video with the initial mask to ensure it matches the reference expression. If a mismatch is found, the check agent corrects the referred object prompt based on the intermediate segmentation results and passes the revised referred object prompt back to the segment agent, which generates the revised mask. 

Unlike previous work \cite{TGS}, which first grounds and then performs segmentation, our RLS directly performs segmentation based on object text prompts, thus avoiding errors caused by inaccurate localization. Moreover, segmentation guided by the initially referred object prompt establishes the foundation for the initial segmentation quality, while reflective learning effectively corrects errors and optimizes the prompt, thereby overcoming performance bottlenecks.

\section{Experiment}

\subsection{Experimental Settings}

\textbf{Dataset.} We evaluate our MAR3 on the Ref-AVSBench dataset \cite{EMCC}, which contains 4,000 audible videos with pixel-level annotations and corresponding expressions, covering 51 object classes. The dataset is divided into 2,908 videos for training, 276 videos for validation, and 818 videos for testing. The test set is further divided into three subsets: (1) Seen set: composed of the 39 categories that appear in training. (2) Unseen set: contains 13 additional categories not seen during the training stage. (3) Null set: refers to expressions that do not correspond to any object in the videos and are excluded from the evaluation in this paper.

\noindent
\textbf{Evaluation Metrics.} Following previous work \cite{EMCC}, we use the Jaccard index (\( \mathcal{J} \)) and the F-score (\( \mathcal{F} \)) as the main evaluation metrics. The Jaccard index (\( \mathcal{J} \)) measures the ratio of intersection to union between predicted and ground truth masks. A higher value indicates better overlap and segmentation performance. The F-score (\( \mathcal{F} \)) is the harmonic mean of precision and recall, which balances accuracy and coverage in predicting positive samples. A higher F-score shows better overall classification performance. Additionally, we report the average of these two metrics (\( \mathcal{J} \, \text{\&} \, \mathcal{F} \)) for a more comprehensive evaluation.

\noindent
\textbf{Implementation Details.} The Frames Per Second of the video is set to 1, and the audio sampling rate is set to 22,050 Hz. In the Consensus Multimodal Recognition mechanism, for MCA Agent \#1, we use Qwen3-30B-A3B-Instruct \cite{qwen3technicalreport}; for MCA Agent \#2, we use GLM4-32B \cite{glm2024chatglm}; and for MCA Agent \#3 and MCA Agent \#4, we use Deepseek-R1-32B \cite{guo2025deepseek}. In the Collaborative Object Reasoning mechanism, for the Visual Agent, Audio Agent, and Audio-Visual Agent, we use Qwen3-Omni-30B-A3B-Instruct \cite{Qwen3-Omni}. In the Reflective Learning Segmentation, for the Check Agent, we use Qwen3-Omni-30B-A3B-Instruct \cite{Qwen3-Omni}, and for the Segment Agent, we use SAM3 \cite{SAM3}. We limit the maximum iterative segmentation to 2 steps. All experiments are conducted on an A100 GPU. Details on the prompts of each stage can be found in Appendix A of the supplementary material.

\subsection{Main Results}

We perform a performance comparison of our proposed MAR3 with previous SOTA methods on the standard Ref-AVSBench dataset. The results of all compared baselines are obtained from the original papers or their official implementations. As shown in Table \ref{tab:comparison}, our MAR3 achieves state-of-the-art performance compared to all these methods in most key metrics. On the \textit{Mix (S+U)} and \textit{Seen} subsets, our MAR3 outperforms the second-best method, TGS-Agent \cite{TGS}, by 3.4\% and 7.4\% in $\mathcal{J} \& \mathcal{F}$, respectively. 
We attribute this substantial performance gain to our core design, \ie unlike previous methods that rely solely on multimodal fusion \cite{EMCC,TSAM-CVPR,SAM2LOVE} or depend heavily on the quality of instruction tuning datasets \cite{Crab,TGS}, MAR3 introduces a training-free multi-agent workflow, which successfully decomposes the complex Ref-AVS task in dynamic audio-visual scenarios into three sub-tasks: deep multimodal cues understanding, reliable object reasoning, and precise segmentation mask prediction. 
Additionally, without relying on training resources, our method achieves comparable performance to TGS-Agent on the \textit{Unseen} subset, with only a 0.5\% difference in $\mathcal{J} \& \mathcal{F}$, which further demonstrates the effectiveness of our training-free MAR3.

\begin{table}[!t]
    \centering

    \caption{Ablation study on each component of our proposed MAR3. 
    CMR denotes the Consensus Multimodal Recognition mechanism, where Delphi refers to the sociological Delphi theory in CMR. COR and RLS represent the Collaborative Object Reasoning mechanism and Reflective Learning Segmentation mechanism, respectively.
    }

    \setlength{\tabcolsep}{3.5pt}

    \resizebox{\linewidth}{!}{
        \begin{tabular}{lcccccc}
            \toprule
            Method & \multicolumn{2}{c}{Mix (S+U) $\uparrow$} & \multicolumn{2}{c}{Seen $\uparrow$} & \multicolumn{2}{c}{Unseen $\uparrow$} \\
            \cmidrule(lr){2-3} \cmidrule(lr){4-5} \cmidrule(lr){6-7}
            & $\mathcal{J}$ & $\mathcal{F}$ & $\mathcal{J}$ & $\mathcal{F}$ & $\mathcal{J}$ & $\mathcal{F}$ \\
            \midrule
            w/o CMR           & 60.8 & 70.9 & 53.0 & 64.8 & 68.5 & 76.9 \\
            w/o Delphi in CMR & 61.5 & 71.9 &  55.2  &  67.2   &  67.8  &  76.5  \\
            w/o COR  & 62.8 & 72.9 & 55.9 & 67.9 & 69.7 & 77.9 \\
            w/o  RLS  & 60.4 & 70.5 & 53.0 & 64.7 & 67.8 & 76.3 \\
            Our MAR3          & \textbf{64.1} & \textbf{74.2} & \textbf{56.1}   & \textbf{68.2}   & \textbf{72.1}   & \textbf{80.2}   \\
            \bottomrule
        \end{tabular}
    }

    \label{ablation_1}
\end{table}

\subsection{Ablation Study}

\textbf{Effectiveness of Consensus Multimodal Recognition.}
We conduct ablation studies to verify the effectiveness of our Consensus Multimodal Recognition (CMR) mechanism, as shown in Table \ref{ablation_1}. First, completely removing the CMR module (w/o CMR) leads to a significant decrease in performance (\eg from 64.1\% to 60.8\% in $\mathcal{J}$ and from 74.2\% to 70.9\% in $\mathcal{F}$ on \textit{Mix (S+U)} set), demonstrating the necessity of explicitly recognizing the difficulty of reference expressions and the dominant modality of multimodal cues before executing Ref-AVS. Furthermore, we validate the impact of the Delphi theory in CMR. Relying solely on single-agent analysis, without the multi-agent consensus mechanism (w/o Delphi in CMR), leads to inferior performance compared to the full model (\eg 61.5\% \vs 64.1\% in $\mathcal{J}$  and 71.9\% \vs 74.2\% in $\mathcal{F}$ on \textit{Mix (S+U)} set). These results indicate that single-agent analysis is prone to potential biases, while introducing a Delphi-based consensus mechanism among multiple MCA agents effectively corrects errors through peer interaction and collective decision-making, thus achieving a deeper and more robust multimodal understanding.

\begin{figure*}[t!] 
    \centering
    \includegraphics[width=0.95\linewidth]{fig/vis_all_v2.pdf}
    \caption{The visualization results of our MAR3, along with the second-best method TGS-Agent. Additional qualitative visualizations and comparative examples can be found in Appendix B of the supplemental material.}
    \label{visualtation}
\end{figure*}

\noindent
\textbf{Effectiveness of Collaborative Object Reasoning.} To verify the effectiveness of the multimodal collaboration strategy within the Collaborative Object Reasoning (COR) mechanism, we conduct ablation studies and the results are shown in the 3rd and 5th rows of Table \ref{ablation_1}. We compare the full model with an ablation model that removes the collaboration mechanism and relies solely on the dominant modality agent for independent reasoning (w/o COR). The results show that the full model achieves an improvement of 1.3\%  (from 62.8\% to 64.1\% in $\mathcal{J}$ and from 72.9\% to 74.2\% in $\mathcal{F}$) on \textit{Mix (S+U)} set compared to the ablation model. This clearly demonstrates the innovative advantage of the COR mechanism, \ie relying solely on a single dominant modality is prone to misjudgment in moderate- or high-difficulty scenarios. By incorporating candidate object lists and reason provided by auxiliary-modality agents, COR effectively overcomes the limitations of single-modality perception, thus leading to more reliable object reasoning.

\begin{table}[!t]
    \centering
    \caption{Ablation study on the maximum number of Reflective-Learning steps in the RLS mechanism.}
    
    \renewcommand{\arraystretch}{1.2}
    \setlength{\tabcolsep}{6pt}
    
    \resizebox{\linewidth}{!}{
        \begin{tabular}{c cccccc} 
            \toprule
            \multirow{2.5}{*}{\shortstack{Reflective-Learning\\$\times N$}} & \multicolumn{2}{c}{Mix (S+U) $\uparrow$} & \multicolumn{2}{c}{Seen $\uparrow$} & \multicolumn{2}{c}{Unseen $\uparrow$} \\
            \cmidrule(lr){2-3} \cmidrule(lr){4-5} \cmidrule(lr){6-7}
            & $\mathcal{J}$ & $\mathcal{F}$ & $\mathcal{J}$ & $\mathcal{F}$ & $\mathcal{J}$ & $\mathcal{F}$ \\
            \midrule
            0 & 60.4 & 70.5 & 53.0 & 64.7 & 67.8 & 76.3 \\
            1 & 63.3 & 73.4 & 55.9 & 67.9 & 70.7 & 78.8 \\
            2 & \textbf{64.1} & \textbf{74.2} & \textbf{56.1} & \textbf{68.2} & \textbf{72.1} & \textbf{80.2} \\
            3 & 63.5 & 73.6 & 55.8 & 67.8 & 71.2 & 79.3 \\
            \bottomrule
        \end{tabular}
    }   
    \label{tab:ablation_RLS_2}
\end{table}

\begin{figure*}[t]
    \centering
    \begin{subfigure}[b]{0.3\textwidth}
        \centering
        \includegraphics[width=\linewidth]{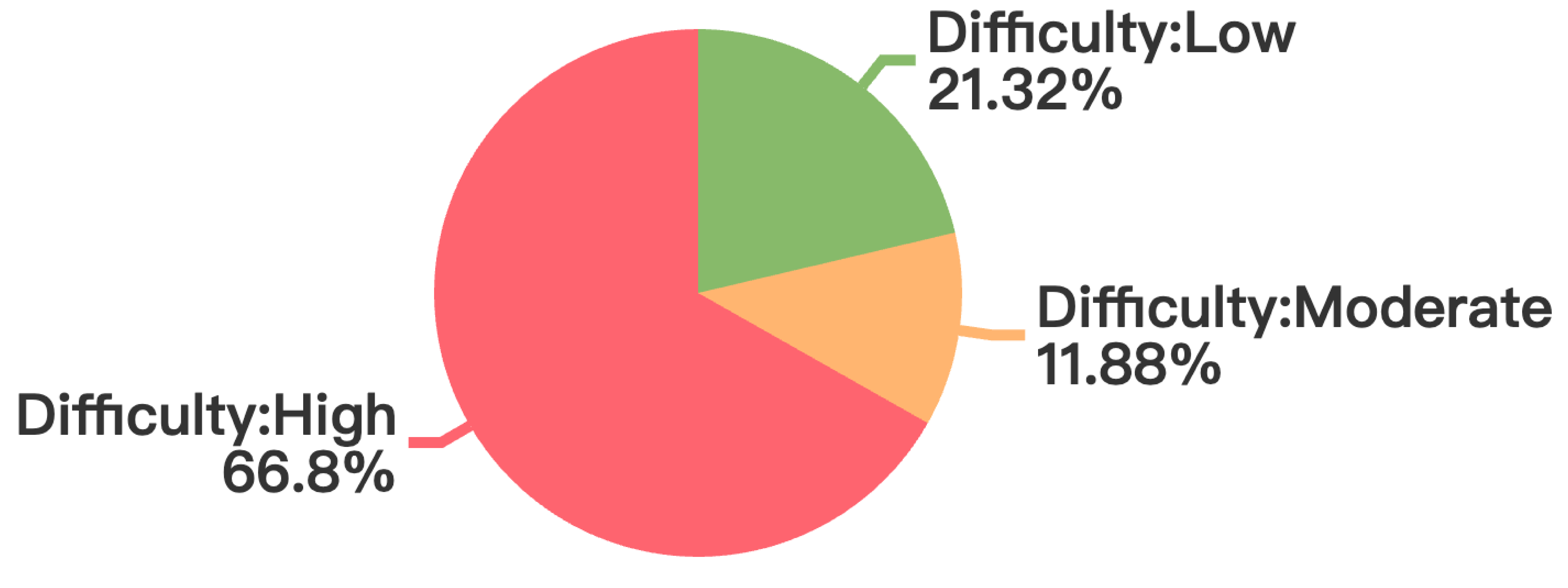}
        \caption{(S+U)}
        \label{fig:diff_test_s_u}
    \end{subfigure}
    \hfill
    \begin{subfigure}[b]{0.3\textwidth}
        \centering
        \includegraphics[width=\linewidth]{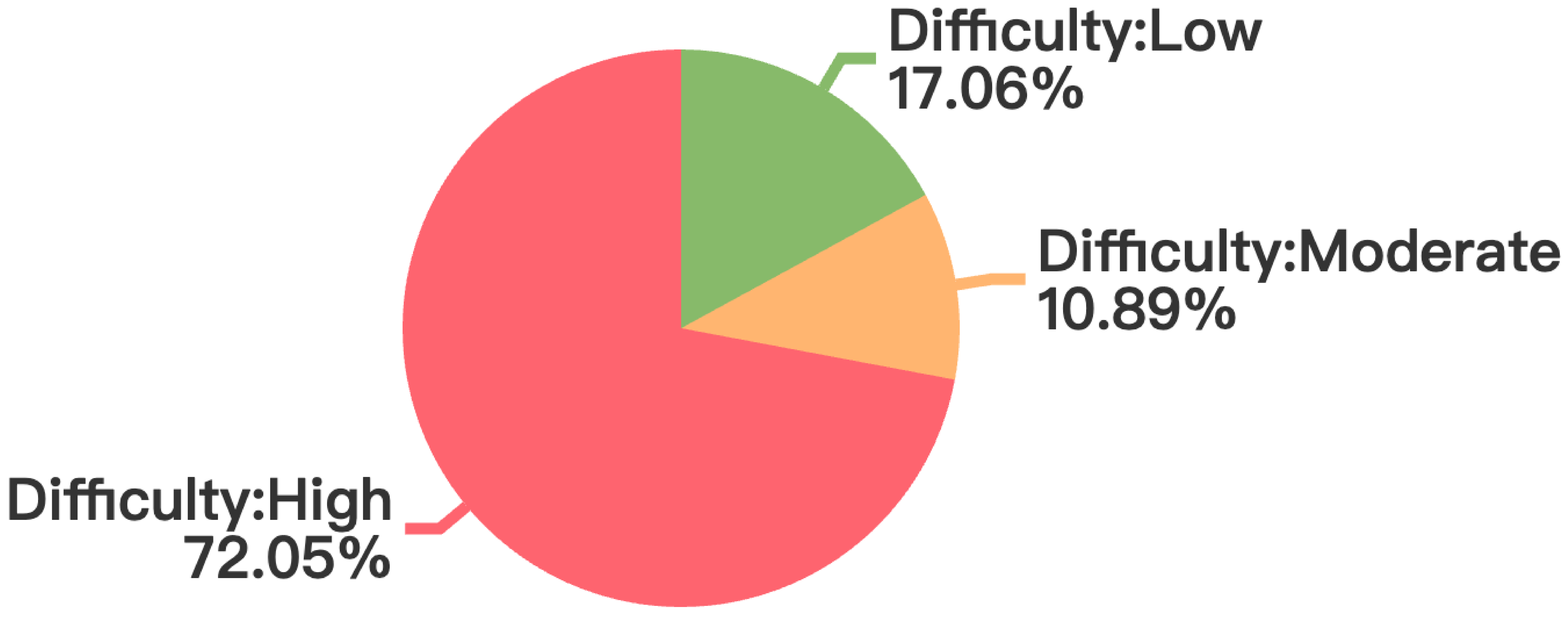}
        \caption{Seen}
        \label{fig:diff_seen}
    \end{subfigure}
    \hfill
    \begin{subfigure}[b]{0.3\textwidth}
        \centering
        \includegraphics[width=\linewidth]{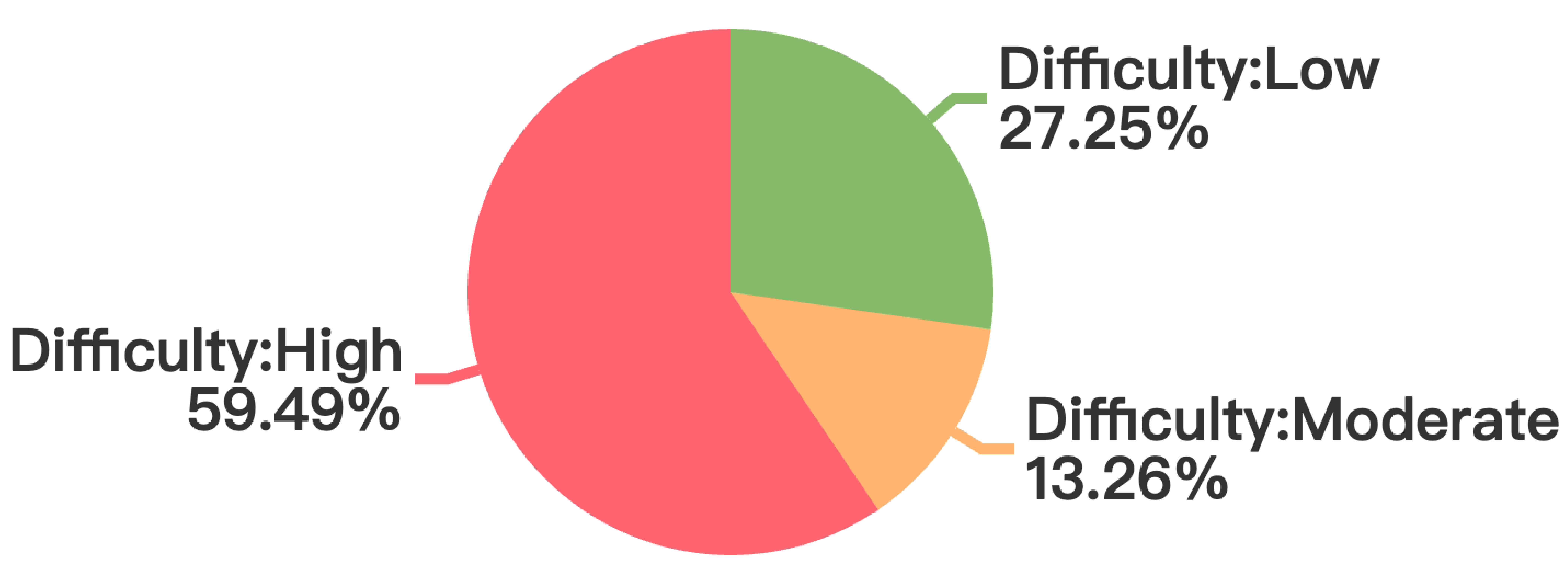}
        \caption{Unseen}
        \label{fig:diff_unseen}
    \end{subfigure}

    \caption{
Difficulty proportions of reference expressions on the Ref-AVSBench test set, identified by our Consensus Multimodal Recognition (CMR) mechanism based on the modality-dominant difficulty rule.
}
    \label{fig:diff_all}
\end{figure*}

\begin{figure*}[t]
    \centering
    \begin{subfigure}[b]{0.3\textwidth}
        \centering
        \includegraphics[width=\linewidth]{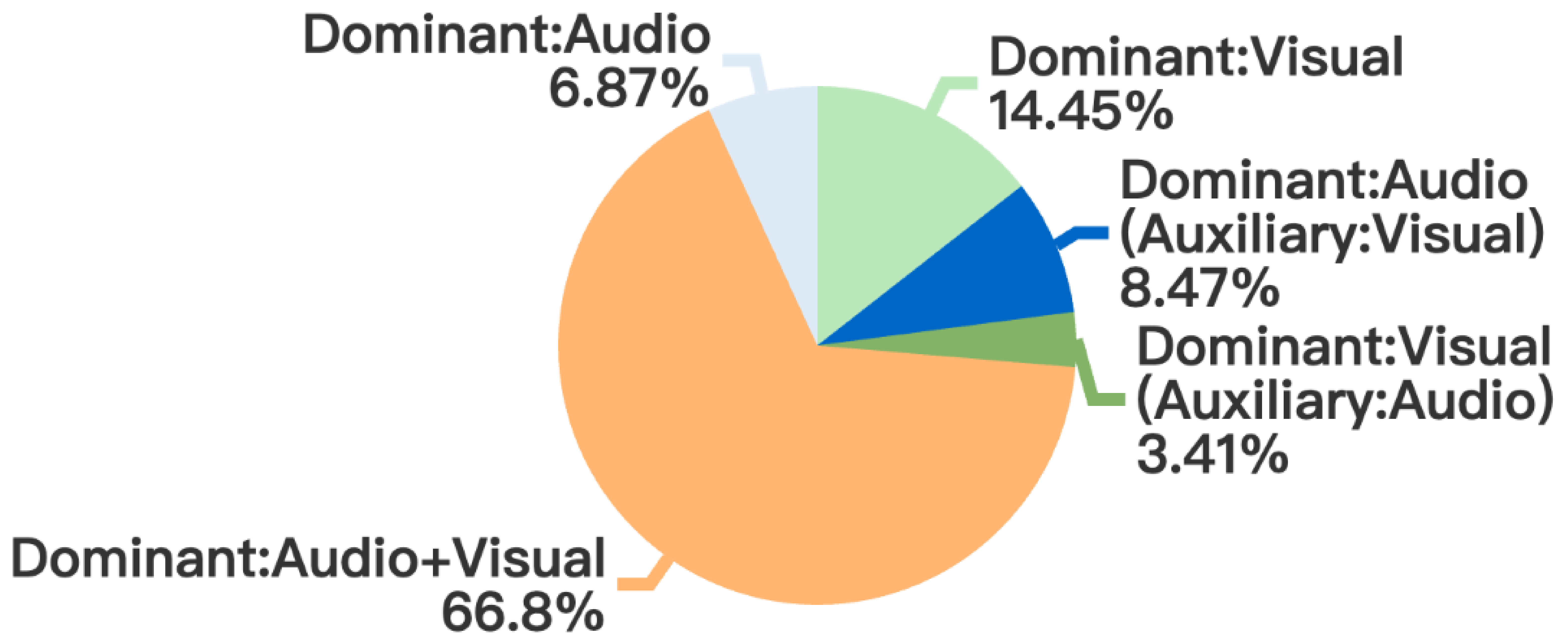}
        \caption{(S+U)}
        \label{fig:modal_test_s_u}
    \end{subfigure}
    \hfill
    \begin{subfigure}[b]{0.3\textwidth}
        \centering
        \includegraphics[width=\linewidth]{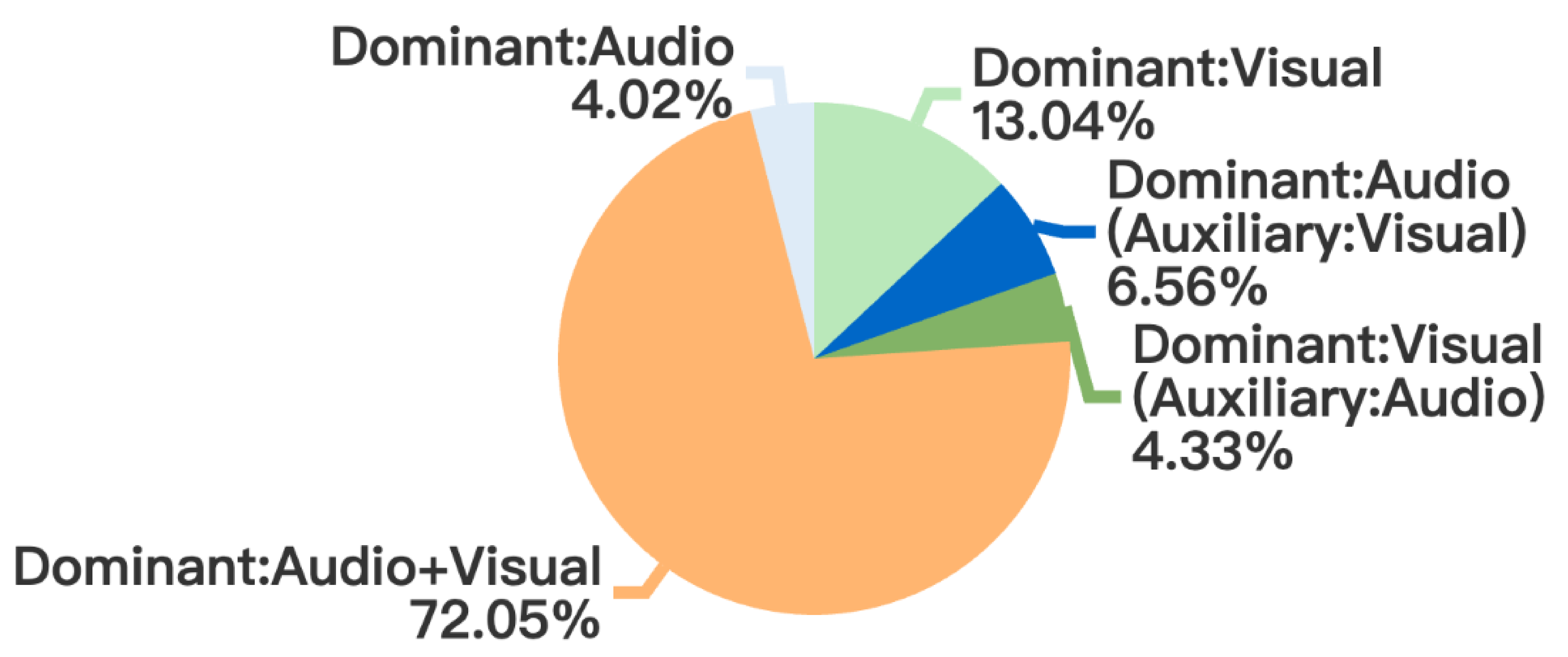}
        \caption{Seen}
        \label{fig:mod_seen}
    \end{subfigure}
    \hfill
    \begin{subfigure}[b]{0.3\textwidth}
        \centering
        \includegraphics[width=\linewidth]{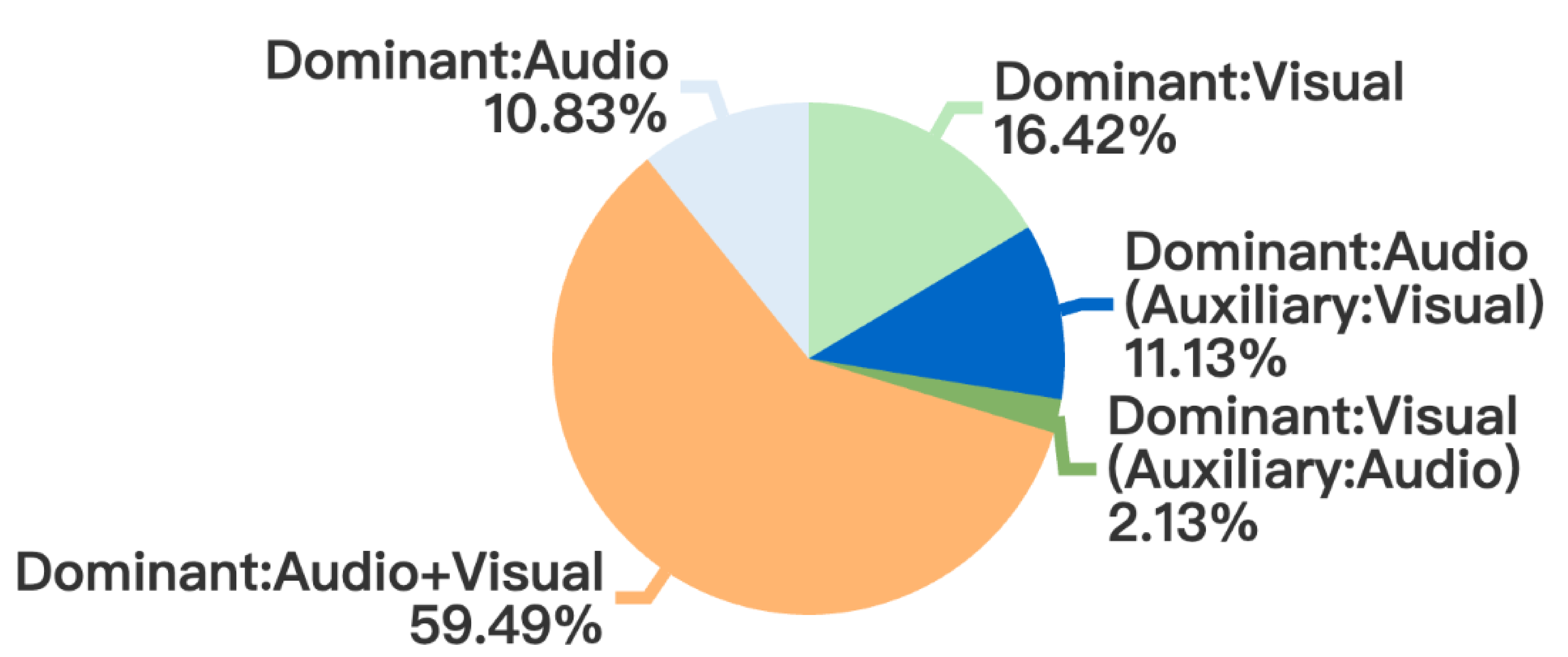}
        \caption{Unseen}
        \label{fig:mod_unseen}
    \end{subfigure}

    \caption{
Dominant modality proportions of reference expressions on the Ref-AVSBench test set, identified by our Consensus Multimodal Recognition (CMR) mechanism based on the modality-dominant difficulty rule.
}
    \label{fig:modality_all}
\end{figure*}

\noindent
\textbf{Effectiveness of Reflective Learning Segmentation.} To validate the effectiveness of the Reflective Learning Segmentation (RLS) mechanism, we conduct ablation experiments, as shown in the 4th and 5th rows of Table \ref{ablation_1}.
We can see that removal of RLS (w/o RLS) leads to a significant performance drop, $\mathcal{J}$ decreases from 64.1\% to 60.4\% and $\mathcal{F}$ decreases from 74.2\% to 70.5\% on \textit{Mix (S+U)} set, which indicates that the iterative referred object prompt revised by the check agent is crucial for overcoming the single-stage segmentation bottleneck and improving mask prediction accuracy. 
Furthermore, we validate the effect of the number of reflective learning iterations $N$ in RLS. As shown in Table \ref{tab:ablation_RLS_2}, performance first improves and then degrades as $N$ increases, achieving the best trade-off at $N=2$. This suggests that moderate reflection effectively corrects initial inference biases, whereas excessive iterations (\eg $N=3$) may over-correct the prompt or introduce unnecessary noise, resulting in a slight performance decline. 
Overall, these results demonstrate that our reflective learning mitigates errors in single-stage segmentation and yields accuracy and robustness gains with only a few iterations ($N=2$).

\subsection{Difficulty and Dominant Modality Proportions of Expressions}

\textbf{Difficulty Proportions.} To gain a deeper understanding of the difficulty of reference expressions in the Ref-AVS task, we report the proportions of different difficulty levels in the Ref-AVSBench test set as recognized by our proposed CMR, as shown in Figure \ref{fig:diff_test_s_u}. The \textit{Mix (S+U)} subset is mainly dominated by high-difficulty samples (66.8\%), while low- and moderate-difficulty cases account for relatively smaller proportions. This indicates that the test set is composed largely of challenging expressions rather than simple ones. Such a distribution of difficulty proportions strongly supports our design motivation of \textquotedblleft analyzing the instruction before executing the task \textquotedblright. Moreover, comparing Figure \ref{fig:diff_seen} and Figure \ref{fig:diff_unseen}, we observe that the Seen subset contains a higher proportion of high-difficulty expressions (72.05\%), whereas the proportion in the Unseen subset is relatively lower (59.49\%). This discrepancy is consistent with the common observation that existing methods \cite{EMCC,TSAM-CVPR,SAM2LOVE,Crab,TGS} often perform better on Unseen samples than on the Seen ones, and it also indirectly demonstrates the effectiveness of our CMR in accurately and reliably recognizing the difficulty of each reference expression.

\noindent
\textbf{Dominant Modality Proportions.} To further analyze the multimodal cues of reference expressions in the Ref-AVS task, we report the proportions of dominant modalities in multimodal cues, as shown in Figure \ref{fig:modal_test_s_u}. \textit{Audio+Visual} co-dominance accounts for the largest portion (66.8\%), which is higher than visual-only dominance (14.45\%) and audio-only dominance (6.87\%). There are also ``dominant + auxiliary'' cases, \ie audio-dominant with visual-auxiliary (8.47\%), and visual-dominant with audio-auxiliary (3.41\%). These results validate the necessity of explicitly recognizing the dominant modality from multimodal cues and provide a basis for the adaptive ``dominant agent + auxiliary agent'' collaborative reasoning in our Collaborative Object Reasoning (COR) module. Moreover, comparing Seen (Figure~\ref{fig:mod_seen}) and Unseen (Figure~\ref{fig:mod_unseen}) subsets, the Seen subset contains a higher proportion of \textit{Audio+Visual} dominance (72.05\%), while the Unseen subset is relatively lower (59.49\%), and the audio-related proportion increases in Unseen (audio-dominant of 10.83\%, and audio-dominant with visual-auxiliary of 11.13\%). This discrepancy suggests that modality dependence varies across subsets and further supports the need for adaptive reasoning paths in COR.

\subsection{Qualitative Analysis}
To provide a more intuitive and comprehensive understanding of the effectiveness of our method, we conduct a qualitative comparison between our proposed MAR3 and the second-best method, TGS-Agent \cite{TGS}. 

\noindent
\textbf{Reliable Reasoning for Referred Objects.} 
As shown in Figure \ref{visualtation} (a), the CMR mechanism based on Delphi theory \cite{linstone1975delphi,Delphi_2,Delphi_3,Delphi_4,Delphi_5,Delphi_6} can effectively recognize both the difficulty level of the reference expression and the dominant modality contained within multimodal cues. Through this process, the COR mechanism is able to adaptively integrate relevant and useful auxiliary information from both audio and visual modalities under a collaborative reasoning paradigm, thereby enabling a more accurate and reliable inference of the initially referred object, while our RLS is further able to yield a more precise segmentation mask prediction.

\noindent
\textbf{Error Correction via Reflective Learning.} 
As illustrated in Figure \ref{visualtation}(b), similar to TGS-Agent which mistakenly identifies “Dog” as the object with the longest sound duration, our method may initially infer an incorrect referent. Nevertheless, the proposed Reflective Learning mechanism allows it to quickly inspect intermediate segmentation outputs and generate a revised referred-object prompt (Hair-dryer) accordingly.
These examples further demonstrate the distinct advantage and effectiveness of our method over the baseline in achieving high-quality reference audio-visual segmentation. Additionally, more qualitative visualizations and detailed comparative examples can be found in Appendix B of the supplementary material.

\section{Conclusion}

In this paper, we propose MAR3, a training-free multi-agent framework that decomposes the complex Ref-AVS task into Recognition, Reasoning, and Reflection, to deeply understand multimodal cues, reliably reason about referred objects, and precisely generate segmentation mask predictions. A Consensus Multimodal Recognition module is designed to explicitly recognize the difficulty of reference expressions and the dominant modality of multimodal cues. An adaptive Collaborative Object Reasoning (COR) strategy is proposed to reliably reason about the referred object by adopting a difficulty-aware collaboration paradigm. Finally, we develop a Reflective Learning Segmentation (RLS) mechanism to ensure precise mask prediction by employing reflective learning between the check agent and the segment agent. Despite its effectiveness, our method has not been specifically designed to handle cases where the expression refers to an object that does not actually exist in the video, but we plan to further explore this special case in our future work. We believe that this work provides important insights and valuable inspiration for the further development of multimodal reasoning and understanding, as well as multi-agent systems. 

\bibliographystyle{ACM-Reference-Format}
\bibliography{sample-base}

@String{Computing = "Computing" }

@String{Computer = "{IEEE} Computer" }

@String{Springer = "Springer-Verlag" }

@ArtifactSoftware{R,
    title = {R: A Language and Environment for Statistical Computing},
    author = {{R Core Team}},
    organization = {R Foundation for Statistical Computing},
    address = {Vienna, Austria},
    year = {2019},
    url = {https://www.R-project.org/},
}

@inproceedings{EMCC,
  title={Ref-avs: Refer and segment objects in audio-visual scenes},
  author={Wang, Yaoting and Sun, Peiwen and Zhou, Dongzhan and Li, Guangyao and Zhang, Honggang and Hu, Di},
  booktitle={European Conference on Computer Vision},
  pages={196--213},
  year={2024},
  organization={Springer}
}

@inproceedings{TSAM-CVPR,
  title={TSAM: Temporal SAM Augmented with Multimodal Prompts for Referring Audio-Visual Segmentation},
  author={Radman, Abduljalil and Laaksonen, Jorma},
  booktitle={Proceedings of the Computer Vision and Pattern Recognition Conference},
  pages={23947--23956},
  year={2025}
}

@inproceedings{SAM2LOVE,
  title={SAM2-LOVE: Segment Anything Model 2 in Language-aided Audio-Visual Scenes},
  author={Wang, Yuji and Xu, Haoran and Liu, Yong and Li, Jiaze and Tang, Yansong},
  booktitle={Proceedings of the Computer Vision and Pattern Recognition Conference},
  pages={28932--28941},
  year={2025}
}

@inproceedings{Crab,
  title={Crab: A unified audio-visual scene understanding model with explicit cooperation},
  author={Du, Henghui and Li, Guangyao and Zhou, Chang and Zhang, Chunjie and Zhao, Alan and Hu, Di},
  booktitle={Proceedings of the Computer Vision and Pattern Recognition Conference},
  pages={18804--18814},
  year={2025}
}

@article{TGS,
  title={Think before you segment: An object-aware reasoning agent for referring audio-visual segmentation},
  author={Zhou, Jinxing and Zhou, Yanghao and Han, Mingfei and Wang, Tong and Chang, Xiaojun and Cholakkal, Hisham and Anwer, Rao Muhammad},
  journal={arXiv preprint arXiv:2508.04418},
  year={2025}
}

@inproceedings{groundingdino,
  title={Grounding dino: Marrying dino with grounded pre-training for open-set object detection},
  author={Liu, Shilong and Zeng, Zhaoyang and Ren, Tianhe and Li, Feng and Zhang, Hao and Yang, Jie and Jiang, Qing and Li, Chunyuan and Yang, Jianwei and Su, Hang and others},
  booktitle={European conference on computer vision},
  pages={38--55},
  year={2024},
  organization={Springer}
}

@article{SAM2,
  title={Sam 2: Segment anything in images and videos},
  author={Ravi, Nikhila and Gabeur, Valentin and Hu, Yuan-Ting and Hu, Ronghang and Ryali, Chaitanya and Ma, Tengyu and Khedr, Haitham and R{\"a}dle, Roman and Rolland, Chloe and Gustafson, Laura and others},
  journal={arXiv preprint arXiv:2408.00714},
  year={2024}
}

@article{multi-agent1,
  title={LLM multi-agent systems: Challenges and open problems},
  author={Han, Shanshan and Zhang, Qifan and Yao, Yuhang and Jin, Weizhao and Xu, Zhaozhuo},
  journal={arXiv preprint arXiv:2402.03578},
  year={2024}
}

@article{multi-agent2,
  title={Multi-agent systems},
  author={Van der Hoek, Wiebe and Wooldridge, Michael},
  journal={Foundations of Artificial Intelligence},
  volume={3},
  pages={887--928},
  year={2008},
  publisher={Elsevier}
}

@inproceedings{MCCD-text-to-image,
  title={MCCD: Multi-Agent Collaboration-based Compositional Diffusion for Complex Text-to-Image Generation},
  author={Li, Mingcheng and Hou, Xiaolu and Liu, Ziyang and Yang, Dingkang and Qian, Ziyun and Chen, Jiawei and Wei, Jinjie and Jiang, Yue and Xu, Qingyao and Zhang, Lihua},
  booktitle={Proceedings of the Computer Vision and Pattern Recognition Conference},
  pages={13263--13272},
  year={2025}
}

@inproceedings{Tree-agent,
  title={Tree-of-Reasoning: Towards Complex Medical Diagnosis via Multi-Agent Reasoning with Evidence Tree},
  author={Peng, Qi and Cui, Jialin and Xie, Jiayuan and Cai, Yi and Li, Qing},
  booktitle={Proceedings of the 33rd ACM International Conference on Multimedia},
  pages={1744--1753},
  year={2025}
}

@article{kim2024mdagents,
  title={Mdagents: An adaptive collaboration of llms for medical decision-making},
  author={Kim, Yubin and Park, Chanwoo and Jeong, Hyewon and Chan, Yik S and Xu, Xuhai and McDuff, Daniel and Lee, Hyeonhoon and Ghassemi, Marzyeh and Breazeal, Cynthia and Park, Hae W},
  journal={Advances in Neural Information Processing Systems},
  volume={37},
  pages={79410--79452},
  year={2024}
}

@inproceedings{chen2025t2i,
  title={T2i-copilot: A training-free multi-agent text-to-image system for enhanced prompt interpretation and interactive generation},
  author={Chen, Chieh-Yun and Shi, Min and Zhang, Gong and Shi, Humphrey},
  booktitle={Proceedings of the IEEE/CVF International Conference on Computer Vision},
  pages={19396--19405},
  year={2025}
}

@inproceedings{audiogen2,
    title = "Orchestrating Audio: Multi-Agent Framework for Long-Video Audio Synthesis",
    author = "Zhang, Yehang  and
      Xu, Xinli  and
      Xu, Xiaojie  and
      Zhang, Doudou  and
      Liu, Li  and
      Chen, Ying-Cong",
    editor = "Christodoulopoulos, Christos  and
      Chakraborty, Tanmoy  and
      Rose, Carolyn  and
      Peng, Violet",
    booktitle = "Proceedings of the 2025 Conference on Empirical Methods in Natural Language Processing",
    month = nov,
    year = "2025",
    address = "Suzhou, China",
    publisher = "Association for Computational Linguistics",
    url = "https://aclanthology.org/2025.emnlp-main.1133/",
    doi = "10.18653/v1/2025.emnlp-main.1133",
    pages = "22267--22282",
    ISBN = "979-8-89176-332-6",

}

@inproceedings{audiogen1,
  title={Audiogenie: A training-free multi-agent framework for diverse multimodality-to-multiaudio generation},
  author={Rong, Yan and Wang, Jinting and Lei, Guangzhi and Yang, Shan and Liu, Li},
  booktitle={Proceedings of the 33rd ACM International Conference on Multimedia},
  pages={8872--8881},
  year={2025}
}

@book{linstone1975delphi,
  title={The delphi method},
  author={Linstone, Harold A and Turoff, Murray and others},
  volume={1975},
  year={1975},
  publisher={Addison-Wesley Reading, MA}
}

@inproceedings{collaborative_1,
    title = "Beyond Frameworks: Unpacking Collaboration Strategies in Multi-Agent Systems",
    author = "Wang, Haochun  and
      Zhao, Sendong  and
      Wang, Jingbo  and
      Qiang, Zewen  and
      Qin, Bing  and
      Liu, Ting",
    editor = "Che, Wanxiang  and
      Nabende, Joyce  and
      Shutova, Ekaterina  and
      Pilehvar, Mohammad Taher",
    booktitle = "Proceedings of the 63rd Annual Meeting of the Association for Computational Linguistics (Volume 1: Long Papers)",
    month = jul,
    year = "2025",
    address = "Vienna, Austria",
    publisher = "Association for Computational Linguistics",
    url = "https://aclanthology.org/2025.acl-long.1037/",
    doi = "10.18653/v1/2025.acl-long.1037",
    pages = "21361--21375",
    ISBN = "979-8-89176-251-0",

}

@inproceedings{collaborative_2,
    title = "Many Heads Are Better Than One: Improved Scientific Idea Generation by A {LLM}-Based Multi-Agent System",
    author = "Su, Haoyang  and
      Chen, Renqi  and
      Tang, Shixiang  and
      Yin, Zhenfei  and
      Zheng, Xinzhe  and
      Li, Jinzhe  and
      Qi, Biqing  and
      Wu, Qi  and
      Li, Hui  and
      Ouyang, Wanli  and
      Torr, Philip  and
      Zhou, Bowen  and
      Dong, Nanqing",
    editor = "Che, Wanxiang  and
      Nabende, Joyce  and
      Shutova, Ekaterina  and
      Pilehvar, Mohammad Taher",
    booktitle = "Proceedings of the 63rd Annual Meeting of the Association for Computational Linguistics (Volume 1: Long Papers)",
    month = jul,
    year = "2025",
    address = "Vienna, Austria",
    publisher = "Association for Computational Linguistics",
    url = "https://aclanthology.org/2025.acl-long.1368/",
    doi = "10.18653/v1/2025.acl-long.1368",
    pages = "28201--28240",
    ISBN = "979-8-89176-251-0",

}

@misc{qwen3technicalreport,
      title={Qwen3 Technical Report}, 
      author={Qwen Team},
      year={2025},
      eprint={2505.09388},
      archivePrefix={arXiv},
      primaryClass={cs.CL},
      url={https://arxiv.org/abs/2505.09388}, 
}

@article{glm2024chatglm,
  title={Chatglm: A family of large language models from glm-130b to glm-4 all tools},
  author={GLM, Team and Zeng, Aohan and Xu, Bin and Wang, Bowen and Zhang, Chenhui and Yin, Da and Zhang, Dan and Rojas, Diego and Feng, Guanyu and Zhao, Hanlin and others},
  journal={arXiv preprint arXiv:2406.12793},
  year={2024}
}

@article{guo2025deepseek,
  title={Deepseek-r1: Incentivizing reasoning capability in llms via reinforcement learning},
  author={Guo, Daya and Yang, Dejian and Zhang, Haowei and Song, Junxiao and Zhang, Ruoyu and Xu, Runxin and Zhu, Qihao and Ma, Shirong and Wang, Peiyi and Bi, Xiao and others},
  journal={arXiv preprint arXiv:2501.12948},
  year={2025}
}

@article{Qwen3-Omni,
  title={Qwen3-Omni Technical Report},
  author={Jin Xu and Zhifang Guo and Hangrui Hu and Yunfei Chu and Xiong Wang and Jinzheng He and Yuxuan Wang and Xian Shi and Ting He and Xinfa Zhu and Yuanjun Lv and Yongqi Wang and Dake Guo and He Wang and Linhan Ma and Pei Zhang and Xinyu Zhang and Hongkun Hao and Zishan Guo and Baosong Yang and Bin Zhang and Ziyang Ma and Xipin Wei and Shuai Bai and Keqin Chen and Xuejing Liu and Peng Wang and Mingkun Yang and Dayiheng Liu and Xingzhang Ren and Bo Zheng and Rui Men and Fan Zhou and Bowen Yu and Jianxin Yang and Le Yu and Jingren Zhou and Junyang Lin},
  journal={arXiv preprint arXiv:2509.17765},
  year={2025}
}

@article{SAM3,
  title={Sam 3: Segment anything with concepts},
  author={Carion, Nicolas and Gustafson, Laura and Hu, Yuan-Ting and Debnath, Shoubhik and Hu, Ronghang and Suris, Didac and Ryali, Chaitanya and Alwala, Kalyan Vasudev and Khedr, Haitham and Huang, Andrew and others},
  journal={arXiv preprint arXiv:2511.16719},
  year={2025}
}

@inproceedings{OISA,
  title={Towards omnimodal expressions and reasoning in referring audio-visual segmentation},
  author={Ying, Kaining and Ding, Henghui and Jie, Guangquan and Jiang, Yu-Gang},
  booktitle={Proceedings of the IEEE/CVF International Conference on Computer Vision},
  pages={22575--22585},
  year={2025}
}

@article{gemini_pro,
  title={Gemini 1.5: Unlocking multimodal understanding across millions of tokens of context},
  author={Team, Gemini and Georgiev, Petko and Lei, Ving Ian and Burnell, Ryan and Bai, Libin and Gulati, Anmol and Tanzer, Garrett and Vincent, Damien and Pan, Zhufeng and Wang, Shibo and others},
  journal={arXiv preprint arXiv:2403.05530},
  year={2024}
}

@inproceedings{multi_agent3,
  title={CompileAgent: Automated real-world repo-level compilation with tool-integrated LLM-based agent system},
  author={Hu, Li and Chen, Guoqiang and Shang, Xiuwei and Cheng, Shaoyin and Wu, Benlong and LiGangyang, LiGangyang and Zhu, Xu and Zhang, Weiming and Yu, Nenghai},
  booktitle={Proceedings of the 63rd Annual Meeting of the Association for Computational Linguistics (Volume 1: Long Papers)},
  pages={2078--2091},
  year={2025}
}

@inproceedings{Genmac,
  title={Genmac: compositional text-to-video generation with multi-agent collaboration},
  author={Huang, Kaiyi and Huang, Yukun and Ning, Xuefei and Lin, Zinan and Wang, Yu and Liu, Xihui},
  booktitle={Proceedings of the AAAI Conference on Artificial Intelligence},
  volume={40},
  number={7},
  pages={5049--5057},
  year={2026}
}

@inproceedings{Hm-rag,
  title={Hm-rag: Hierarchical multi-agent multimodal retrieval augmented generation},
  author={Liu, Pei and Liu, Xin and Yao, Ruoyu and Liu, Junming and Meng, Siyuan and Wang, Ding and Ma, Jun},
  booktitle={Proceedings of the 33rd ACM international conference on multimedia},
  pages={2781--2790},
  year={2025}
}

@inproceedings{text-image-2,
  title={Promptsculptor: Multi-agent based text-to-image prompt optimization},
  author={Xiang, Dawei and Xu, Wenyan and Chu, Kexin and Ding, Tianqi and Shen, Zixu and Zeng, Yiming and Su, Jianchang and Zhang, Wei},
  booktitle={Proceedings of the 2025 Conference on Empirical Methods in Natural Language Processing: System Demonstrations},
  pages={774--786},
  year={2025}
}

@inproceedings{zhou2025mam,
  title={Mam: Modular multi-agent framework for multi-modal medical diagnosis via role-specialized collaboration},
  author={Zhou, Yucheng and Song, Lingran and Shen, Jianbing},
  booktitle={Findings of the Association for Computational Linguistics: ACL 2025},
  pages={25319--25333},
  year={2025}
}

@inproceedings{li2025dialogueagents,
  title={Dialogueagents: A hybrid agent-based speech synthesis framework for multi-party dialogue},
  author={Li, Xiang and Pan, Duyi and Xiao, Hongru and Han, Jiale and Tang, Jing and Ma, Jiabao and Wang, Wei and Cheng, Bo},
  booktitle={2025 IEEE International Conference on Multimedia and Expo (ICME)},
  pages={1--6},
  year={2025},
  organization={IEEE}
}

@inproceedings{Lvagent,
  title={Lvagent: Long video understanding by multi-round dynamical collaboration of mllm agents},
  author={Chen, Boyu and Yue, Zhengrong and Chen, Siran and Wang, Zikang and Liu, Yang and Li, Peng and Wang, Yali},
  booktitle={Proceedings of the IEEE/CVF International Conference on Computer Vision},
  pages={20237--20246},
  year={2025}
}

@inproceedings{wang2024macrec,
  title={Macrec: A multi-agent collaboration framework for recommendation},
  author={Wang, Zhefan and Yu, Yuanqing and Zheng, Wendi and Ma, Weizhi and Zhang, Min},
  booktitle={Proceedings of the 47th International ACM SIGIR Conference on Research and Development in Information Retrieval},
  pages={2760--2764},
  year={2024}
}

@inproceedings{SAM,
  title={Segment anything},
  author={Kirillov, Alexander and Mintun, Eric and Ravi, Nikhila and Mao, Hanzi and Rolland, Chloe and Gustafson, Laura and Xiao, Tete and Whitehead, Spencer and Berg, Alexander C and Lo, Wan-Yen and others},
  booktitle={Proceedings of the IEEE/CVF international conference on computer vision},
  pages={4015--4026},
  year={2023}
}

@inproceedings{ding2025muses,
  title={Muses: 3d-controllable image generation via multi-modal agent collaboration},
  author={Ding, Yanbo and Zhuang, Shaobin and Li, Kunchang and Yue, Zhengrong and Qiao, Yu and Wang, Yali},
  booktitle={Proceedings of the AAAI Conference on Artificial Intelligence},
  volume={39},
  number={3},
  pages={2753--2761},
  year={2025}
}

@inproceedings{yang2025nader,
  title={Nader: Neural architecture design via multi-agent collaboration},
  author={Yang, Zekang and Zeng, Wang and Jin, Sheng and Qian, Chen and Luo, Ping and Liu, Wentao},
  booktitle={Proceedings of the Computer Vision and Pattern Recognition Conference},
  pages={4452--4461},
  year={2025}
}

@inproceedings{ouyang2025nvagent,
  title={nvagent: Automated data visualization from natural language via collaborative agent workflow},
  author={Ouyang, Geliang and Chen, Jingyao and Nie, Zhihe and Gui, Yi and Wan, Yao and Zhang, Hongyu and Chen, Dongping},
  booktitle={Proceedings of the 63rd Annual Meeting of the Association for Computational Linguistics (Volume 1: Long Papers)},
  pages={19534--19567},
  year={2025}
}

@article{hu2024storyagent,
  title={Storyagent: Customized storytelling video generation via multi-agent collaboration},
  author={Hu, Panwen and Jiang, Jin and Chen, Jianqi and Han, Mingfei and Liao, Shengcai and Chang, Xiaojun and Liang, Xiaodan},
  journal={arXiv preprint arXiv:2411.04925},
  year={2024}
}

@article{Delphi_2,
  title={Using expertsopinions through Delphi technique},
  author={Yousuf, Muhammad Imran},
  journal={Practical assessment, research, and evaluation},
  volume={12},
  number={1},
  year={2007},
  publisher={University of Massachusetts Amherst Libraries}
}

@article{Delphi_3,
  title={The design of a policy Delphi},
  author={Turoff, Murray},
  journal={Technological forecasting and social change},
  volume={2},
  number={2},
  pages={149--171},
  year={1970},
  publisher={Elsevier}
}

@article{Delphi_4,
  title={The decision delphi},
  author={Rauch, Wolf},
  journal={Technological forecasting and social change},
  volume={15},
  number={3},
  pages={159--169},
  year={1979},
  publisher={Elsevier}
}

@article{Delphi_5,
  title={Consensus in the Delphi method: what makes a decision change?},
  author={Barrios, Maite and Guilera, Georgina and Nu{\~n}o, Laura and G{\'o}mez-Benito, Juana},
  journal={Technological Forecasting and Social Change},
  volume={163},
  pages={120484},
  year={2021},
  publisher={Elsevier}
}

@article{Delphi_6,
  title={Research pearls: expert consensus based evidence using the Delphi method},
  author={Hohmann, Erik and Cote, Mark P and Brand, Jefferson C},
  journal={Arthroscopy: the journal of arthroscopic \& related surgery},
  volume={34},
  number={12},
  pages={3278--3282},
  year={2018},
  publisher={Elsevier}
}

\end{document}